\documentclass[a4paper,12pt,english,german]{article}

\begin{document}

{\bf  Quantum Correlations Relativity for Continuous Variable
Systems }

\bigskip

{ M. Dugi\' c$^{\dag}$, M. Arsenijevi\' c$^{\dag}$, J. Jekni\'
c-Dugi\' c$^{\ast}$}

$^{\dag}${Department of Physics, Faculty of Science, Kragujevac,
Serbia}

$^{\ast}${Department of Physics, Faculty of Science, Ni\v s,
Serbia}

\bigskip

Email: dugic@open.telekom.rs, fajnman@gmail.com,
jjeknic@pmf.ni.ac.rs

\bigskip

{\bf Abstract: } It is shown that a choice of  degrees of freedom
of a bipartite continuous variable system determines amount of
non-classical correlations (quantified by discord) in the system's
state. Non-classical correlations (that include entanglement as a
special kind of correlations) are ubiquitous for such systems. For
a quantum state, if there are not non-classical correlations
(quantum discord is zero) for one, there are in general
non-classical correlations (quantum discord is non-zero) for
another set of the composite system's degrees of freedom. The
physical relevance of this 'quantum correlations relativity' is
emphasized also in the more general context.

\bigskip

Key words: entanglement, non-classical correlations, quantum
discord, tensor product structure.

PACS: 03.65.Ca    , 03.67.-a ,    03.65.Ud

\pagebreak

{\bf 1. Introduction}

\bigskip

The promise of the quantum information processing is the promise
of the quantum-information resources [1]. To this end,  some
surprising results and observations are possible and even
expectable. The discovery of non-classical (quantum) correlations
not necessarily including entanglement, as quantified by quantum
discord [2, 3], opens a new avenue in quantum information
processing; for recent reviews see  [4, 5, 6]. A search for
quantum information resources and the ways of their operational
use is at the core of the current theoretical and experimental
research [4, 5, 6, 7, 8] (and the references therein).

Entanglement Relativity is a corollary of the universally valid
quantum mechanics that states [9, 10, 11, 12, 13, 14, 15]: for a
composite (e.g. bipartite) system, there is entanglement for at
least one structure (one set of the degrees of freedom) of the
composite system. The structures being mutually related by the
proper (e.g. the linear canonical) transformations of the
composite system's degrees of freedom; paradigmatic are the
composite system's center-of-mass and the "relative (internal)"
degrees of freedom. In practice, it means: if a quantum state is
separable (no entanglement), just change the degrees of freedom
and entanglement will appear [10, 12, 13]. Quantum entanglement is
ubiquitous as a quantum information resource.

In this paper we consider the continuous variable (CV), including
open, quantum systems with an emphasis on their bipartitions.
Based on Entanglement Relativity, we point out relativity, i.e.
structure (degrees of freedom) dependence, of the more general
non-classical correlations quantified by quantum discord. Likewise
entanglement, the more general non-classical (quantum)
correlations are also structure-dependent and ubiquitous in
quantum systems.

So we conclude:  There are non-classical correlations (not
necessarily including entanglement) for practically every quantum
state of the systems relative to some structures.

In Section 2, we briefly outline Entanglement Relativity. In
Section 3 we derive our main result. Section 4 is discussion
placing our considerations  in a more general context and we
conclude in Section 5.

\bigskip

{\bf 2. Entanglement relativity}

\bigskip

We consider a  composite system $C$ that can be decomposed as
$1+2$, or $A+B$. The continuous degrees of freedom of the
subsystems being mutually related by the proper linear canonical
transformations (LCTs) [10, 12, 13, 14]. Then, by a definition,
the $C$'s Hilbert state space, $\mathcal{H}_C$, can be factorized
as $\mathcal{H}_1 \otimes \mathcal{H}_2$
 as well as $\mathcal{H}_A \otimes \mathcal{H}_B$;
$\mathcal{H}_1 \otimes \mathcal{H}_2 = \mathcal{H}_C =
\mathcal{H}_A \otimes \mathcal{H}_B$.  The two decompositions of
$C$ represent the two different "structures" of $C$. There is a
one-to-one relation between the "structure (decomposition)" and
the composite system's state space factorization. In general,
every subsystem ($1$ or $2$, or $A$ or $B$) can be of arbitrary
number of the degrees of freedom.

Entanglement Relativity establishes [9, 10, 11, 12, 13, 14, 15]:
if a (pure) state is separable for one structure (one set of
degrees of freedom), it typically becomes entangled for another
structure of the composite system:

\begin{equation}
\vert i \rangle_1 \vert j \rangle_2 = \sum_{\alpha, \beta}
C^{ij}_{\alpha\beta} \vert \alpha \rangle_A \vert \beta \rangle_B,
\quad C_{\alpha\beta} \neq  a_{\alpha} b_{\beta},
\sum_{\alpha,\beta}  C^{ij}_{\alpha\beta}  C^{i'j'}_{\alpha\beta}
= \delta_{ii'} \delta_{jj'}.
\end{equation}

The proof of Eq.(1) easily follows, cf. e.g. Ref. [13], with the
use of the covariance function $C_f = \langle \hat A_A \otimes
\hat B_B \rangle - \langle \hat A_A \rangle \langle \hat B_B
\rangle$. Given a separable state for $1+2$ structure, e.g. $\vert
i \rangle_1 \vert j \rangle_2$, and for a pair of the observables
$\hat A_A, \hat B_B$ of the subsystems $A$ and $B$ respectively,
the condition $C_f = 0$  is necessary in order the  state  can
bear separable (tensor-product) form {\it also} for $A+B$
structure. The condition $C_f = 0$ is
 not yet sufficient for the separability.

 The exceptions from Eq.(1) are known--there are the states of the
 separable form for both structures [13, 16].
Nevertheless, as the number of entangled states is by far larger
than the number of the tensor-product states, the number of states
satisfying Eq.(1) is incomparably larger than the number of states
not satisfying Eq.(1). This is the reason in practice these
exceptional cases are usually neglected.

So, one may say: there is entanglement for all pure quantum states
relative to certain structure--i.e. entanglement is a matter of a
composite-system's structure and {\it is} present for one
structure, at least. Certainly, for a fixed structure, a pure
quantum state is either separable (tensor-product) or entangled.

Now, it is natural to wonder if something analogous can be
expected for the mixed quantum states. Of course, now the question
refers not only to quantum entanglement, but also to the more
general quantum, i.e. non-classical, correlations as quantified by
quantum discord  [2,3]. In the next section we provide a
generalization of Entanglement Relativity that directly includes
open quantum systems: [likewise entanglement itself] the more
general non-classical correlations are subject to relativity, i.e.
are ubiquitous regarding the continuous variable bipartite
systems.

\bigskip

{\bf 3. Quantum correlations relativity}

\bigskip

Quantum Discord is a common term for different measures of
non-classical correlations in composite (e.g. bipartite) quantum
systems [1-8]. Historically the first and probably the best known
is the so-called "one-way" discord. For completeness,  we give the
formal definitions of both one-way and two-way discord.

One-way quantum discord for the $S+S'$ system, $D_S = I(S:S') -
J_S \ge 0$, and von Neumann entropy of a state $\rho$,
$\mathcal{S} = - tr \rho\ln\rho$. Both the total mutual
information, $I(S:S') = \mathcal{S}(S) + \mathcal{S}(S') -
\mathcal{S}(S,S')$, and the classical correlations, $J_S =
\mathcal{S}(S) - \inf_{\{\Pi_{S'i}\}}\sum_i \vert c_i \vert^2
\mathcal{S}(\rho_S\vert_{\Pi_{S'i}})$--where
$\rho_S\vert_{\Pi_{S'i}} = I_S \otimes \Pi_{S'i} \rho I_S \otimes
\Pi_{S'i}$ is the state remaining after a selective quantum
measurement defined by the projectors $\Pi_{S'i}$--are
non-negative. Two-way discord, $D = max\{D_S, D_{S'}\}$, where
$D_{S'}$ is one-way discord referring to the $S'$ system.

In this section, we proceed with considering the bipartite
structures of a composite system $C$ with arbitrary number of the
continuous degrees of freedom.

\bigskip

{\it 3.1 Quantum correlations relativity}

\bigskip

Let us consider the $1+2$ structure of the composite system $C$.
The one-way discord $D_1$ equals zero if and only if the composite
system's state, $\hat \rho_C$, is of the form [4, 5, 6]:

\begin{equation}
\hat \rho_C = \sum_k p_k \vert k \rangle_1\langle k \vert \otimes
\hat \rho_{2k}, \sum_k p_k = 1
\end{equation}

\noindent and analogously for the other $D_2$ discord. It is easy
to prove for $\hat \rho_C$ in Eq.(2) that both  the one-way
discord $D_2$ (in general) and the covariance function $C_f$
(Section 2) are nonzero. Of course, the later is a consequence of
the classical correlations in $1+2$ decomposition. Introducing
$\hat \rho_{2k} = \sum_l \omega^k_l \vert \chi^k_l
\rangle_2\langle \chi^k_l \vert$, $\sum_l \omega^k_l = 1,
\forall{k}$ into Eq.(2) gives

\begin{equation}
\hat \rho_C = \sum_{k,l} p_k \omega^k_l \vert k \rangle_1\langle k
\vert \otimes \vert \chi^k_l \rangle_2\langle \chi^k_l \vert.
\end{equation}

Now, with the use of entanglement relativity Eq.(1), we introduce
the alternate structure, $A+B$, into considerations. Let us first,
in disagreement with Eq.(1), the states for both structures are
tensor-product, $\vert k \rangle_1 \vert \chi^k_l\rangle_2 = \vert
k\rangle_A \vert \phi^k_l\rangle_B, \forall{k}$. Then the form
Eq.(2) of the composite system's state is valid also for the $A+B$
structure:

\begin{equation}
\hat \rho_C = \sum_k p_k \vert k \rangle_A\langle k \vert \otimes
\hat \rho_{Bk}, \sum_k p_k = 1,
\end{equation}

\noindent i.e. the one-way discord $D_A = 0$ (in general, $D_B
\neq 0$) while again $C_f \neq 0$.

However, Entanglement Relativity Eq.(1) leads to the following
form of Eq. (3):

\begin{equation}
\hat \rho_C = \sum_{k,l,\alpha, \beta, \beta'} p_k \omega^k_l
C^{kl}_{\alpha\beta} C^{kl\ast}_{\alpha \beta'} \vert \alpha
\rangle_A\langle \alpha \vert \otimes \vert \beta \rangle_B\langle
\beta' \vert + \sum_{k,l,\alpha\neq \alpha', \beta, \beta'} p_k
\omega^k_l C^{kl}_{\alpha\beta} C^{kl\ast}_{\alpha' \beta'} \vert
\alpha \rangle_A\langle \alpha' \vert \otimes \vert \beta
\rangle_B\langle \beta' \vert.
\end{equation}

Clearly from Eq.(5): in order for some basis $\vert
\alpha\rangle_A$ to provide $D_A = 0$, the second term on the rhs
of Eq.(5) must vanish. Due to the linear independence of the terms
$\vert \alpha \rangle_A\langle \alpha' \vert \otimes \vert \beta
\rangle_B\langle \beta' \vert$, this can happen only if:

\begin{equation}
\sum_{k,l} p_k \omega^k_l C^{kl}_{\alpha\beta} C^{kl\ast}_{\alpha'
\beta'} = 0, \forall{\alpha \neq \alpha'}, \forall{\beta, \beta'}.
\end{equation}

Eq. (6) is actually a set of the simultaneously satisfied
equalities. The number of the equalities is equal to the number of
the combinations for the indices $\alpha \neq \alpha'$ and $\beta$
and $\beta'$. Clearly, for the continuous variable systems, this
number is infinite. On the other hand, there is some freedom in
choice of $p_k$ and $\omega_l^k$, as well as the normalization
conditions (cf. Eq.(1)), $\sum_{\alpha,\beta} C^{kl}_{\alpha
\beta} C^{k'l'\ast}_{\alpha \beta} = \delta_{kk'} \delta_{ll'}$.
This may reduce the number of the expressions Eq.(6) that should
be simultaneously satisfied. So, we cannot exclude that there
exist some states of the form Eq.(4) for both structures, $1+2$
and $A+B$. Nevertheless, for every combination of the coefficients
$C_{\alpha\beta}^{kl}$ satisfying Eq.(6), there is the infinite
number of variations of the coefficients $p_k$ and $ \omega^k_l$
that do not satisfy Eq.(6). In practice, it means one may forget
about the states fulfilling Eq.(6), i.e. bearing zero one-way
discord for a pair of structures, $1+2$ and $A+B$.

Regarding the two-way discord ($D$) for the $1+2$ structure, the
condition $D = 0$ (i.e. $D_1 = 0 = D_2$) can be satisfied if and
only if the composite system's state can be written as [4, 5, 6]:

\begin{equation}
\hat \rho_C = \sum_{kl} p_{kl} \vert k\rangle_1\langle k \vert
\otimes \vert l \rangle_2\langle l \vert, \sum_{k,l} p_{kl} =1.
\end{equation}

\noindent Such states are now commonly termed classical-classical
(CC) states. The presence of only classical correlations in CC
states [3] is revealed by $C_f \neq 0$, which is straightforward
to show for Eq.(7). Again, assuming non-validity of Eq. (1), i.e.
assuming  equality $\vert k \rangle_1 \vert l\rangle_2 = \vert
k\rangle_A \vert l\rangle_B, \forall{k, l}$, gives directly rise
to the form Eq.(7) also for the structure $A+B$, i.e. $D_A =
 0 = D_B$.

However, substitution of Eq.(1) for $\vert k \rangle_1 \vert
l\rangle_2$ into Eq.(7) gives rise to:

\begin{equation}
\hat \rho_C = \sum_{k, l, \alpha, \beta} p_{kl} \vert
C^{kl}_{\alpha\beta} \vert^2 \vert \alpha \rangle_A\langle \alpha
\vert \otimes \vert \beta \rangle_B\langle \beta \vert + \sum_{k,
l, \alpha\neq \alpha', \beta \neq \beta'} p_{kl}
C^{kl}_{\alpha\beta} C^{kl\ast}_{\alpha' \beta'} \vert \alpha
\rangle_A\langle \alpha' \vert \otimes \vert \beta
\rangle_B\langle \beta' \vert.
\end{equation}

In order Eq.(8) to be a CC state--i.e. to be of the form
Eq.(7)--also for the $A+B$ structure, the following conditions
should be satisfied:

\begin{equation}
\sum_{k,l} p_{kl} C^{kl}_{\alpha\beta} C^{kl\ast}_{\alpha' \beta'}
= 0, \forall{\alpha \neq \alpha'}, \forall{\beta \neq \beta'}.
\end{equation}

The coefficients $p_{k}\omega_l^k$ in Eq.(6) are variants of the
more general form $p_{kl}$ appearing in Eq.(9). In other words:
Eq. (6) is a variant of the more general and more stringent
equation Eq.(9). So, likewise for Eq.(6), the number of states
satisfying Eq.(9) is negligible compared to the number of states
not fulfilling Eq.(9). Likewise for the one-way discord, one may
practically ignore the possible existence of the states bearing
zero two-way discord for a pair of structures, $1+2$ and $A+B$.

Thereby, the number of the possible solutions to Eq.(6) as well as
to Eq. (9) is by far negligible compared to the number of the
states not fulfilling Eq.(6) i.e. Eq.(9). This observation clearly
emphasizes Quantum Correlations Relativity (QCR) as a new
corollary of the universally valid quantum mechanics:
non-classical correlations are ubiquitous for quantum systems. If
there are not correlations (quantum discord is zero) for one
structure (for one set of the composite system's degrees of
freedom), there are certainly non-classical correlations (non-zero
discord) for an alternative structure (for an alternative set of
the degrees of freedom) of the composite system. Formally, QCR is
presented by the following equality:

\begin{eqnarray}
&\nonumber& \sum_k p_k \vert k\rangle_1\langle k \vert \otimes
\vert k \rangle_2\langle k \vert  = \sum_{k, l, \alpha, \beta}
p_{kl} \vert C^{kl}_{\alpha\beta} \vert^2 \vert \alpha
\rangle_A\langle \alpha \vert \otimes \vert \beta \rangle_B\langle
\beta \vert\\&&
 + \sum_{k, l, \alpha\neq \alpha', \beta
\neq \beta'} p_{kl} C^{kl}_{\alpha\beta} C^{kl\ast}_{\alpha'
\beta'} \vert \alpha \rangle_A\langle \alpha' \vert \otimes \vert
\beta \rangle_B\langle \beta' \vert,
\end{eqnarray}

\noindent and analogously for the one-way discord(s).

So one can say: "non-classical correlations" is not a
characteristic of a composite {\it system}, but is a
characteristic of a composite system's {\it structure}.

\bigskip

{\it 3.2 Comments}

\bigskip

From the previous section we learn: adding a new structure into
considerations reduces the number of states bearing the form
Eq.(4) for all the structures. Thereby we conclude, although not
rigorously prove: every quantum state bears non-classical
(quantum) correlations, for at least one structure of the
composite system. The lack of the rigorous proof is in intimate
relation to the following more general considerations.

While the variables transformations is a universal physical
method, we have only recently started to understand its importance
in the quantum mechanical context. This may be a consequence of
the fact that the following, easily formulated, task is   {\it
open} in the quantum mechanical formalism:

\smallskip

\noindent ({\bf T}) {\it Starting from the left-hand (right-hand)
side} {\it to obtain the right-hand (left-hand) side of} Eq. (1)
{\it i.e. of} Eq. (10).

\smallskip

Regarding Eq.(10), the task {\bf T}  reads: given a quantum state
$\hat \rho_C$ (e.g. Eq.(7) for the $1+2$ structure) to check if
the state takes the separable form Eq.(7) for the structure $A+B$.
This is the task of the Quantum Separability Problem (QUSEP)
thoroughly investigated in the literature for the
finite-dimensional  systems (see e.g. Gharibian [17] and the
references therein). Solving the equations (6) and (9) is clearly
also an instance of the QUSEP problem, which suggests that the
general solutions to Eqs. (6) and (9) are hardly expectable.

\bigskip

{\bf 4. Discussion}

\bigskip

"Quantum discord" is designed to capture all the kinds of
non-classical correlations, including "entanglement", in quantum
systems. Therefore here formulated Quantum Correlations Relativity
(QCR) generalizes Entanglement Relativity, Section 2.

In analogy with [18], QCR can be expressed as "non-classical
(quantum) correlations for (practically) {\it all} quantum states
relative to some structures". Thus we go beyond the "almost all
quantum states have nonclassical correlations" of Ferraro et al
[18]--this "almost" is lost in our conclusion. In  terms of the
result of Ferraro et al [18], QCR can be readily expressed as
follows: if a
 quantum state falls within the zero-discord set
$\Omega_{\circ}$ for one, it is highly improbable to fall within
the analogous $\Omega_{\circ}$ set for virtually any other
structure of the composite system. While the result of Ferraro et
al [18] refers to Markovian open systems, our finding bears
universality.

It is worth emphasizing: quantum correlations relativity is not a
consequence of the reference-frame change or of the more general
relativistic considerations such as e.g. in [19, 20]. The
degrees-of-freedom transformations implicit to our considerations
cannot be written in a separable form for the unitary operators,
i.e. in the form $U_1\otimes U_2$ for the $1+2$ structure--such
transformations are known to preserve discord [4, 5, 6] (and the
references therein). Interestingly enough, some formally trivial
variables transformations exhibit QCR also for the
finite-dimensional (e.g. qubit) systems.

To illustrate, consider a three-qubit system, $\mathcal{C} =
1+2+3$, and its bipartite structures, $1+S_1$ and $S_2+3$, where
the bipartite systems $S_1=2+3$ and $S_2=1+2$.  As it is well
known from quantum teleportation [21], the $\mathcal{C}$'s state
$\vert \phi\rangle_1 \vert \Phi^{+}\rangle_{S_1}$, where $\vert
\Phi^{+}\rangle_{S_1} = (\vert 0\rangle_2 \vert 0\rangle_3 + \vert
1\rangle_2 \vert 1\rangle_3)/2^{-1/2}$, can be re-written as
$\sum_i \vert \chi_i\rangle_{S_2}\vert i\rangle_3/2$, where the
$S_2$'s states represent the Bell states [1] for the pair $1+2$.
The point is that for the $1+S_1$  structure, the state is
tensor-product and therefore not bearing any correlations between
the $1$ and $S_1$ systems, while there is entanglement in the
$S_2+3$ structure.

In our considerations,  entanglement relativity (Section 2) is
basic to the more general correlations relativity, Section 3. Not
surprisingly, as we use  entanglement relativity for the pure
quantum states, which are the building blocks of the mixed states.
The inverse, however, is not in general correct: entanglement
relativity (even for the mixed states) does not in general follow
from the more-general-correlations relativity.

To see this, let us consider a structure $1+2$ for which $D_1  =
0$. Then the quantum state is of the form $\rho = \sum_i p_i \vert
i\rangle_1\langle i \vert \otimes \rho_{2i}$; $\sum_i p_i = 1$.
Now, the correlation relativity suggests there is a structure
$A+B$ for which  $\vert i\rangle_1\langle i \vert \otimes
\rho_{2i} = \sum_j \omega^i_j \rho_{Aj} \otimes \rho_{Bj}$;
$\sum_j \omega^i_j = 1, \forall{i}$. Now, by substituting the
later into the initial form for $\rho$ gives rise to $\rho =
\sum_j \lambda_j \rho_{Aj} \otimes \rho_{Bj}$; $\lambda_j \equiv
\sum_i \omega^i_j$, and $\sum_j \lambda_j = 1$. Certainly, for
this one easily obtains $D_A \neq 0, D_{B} \neq 0$,  $C_f \neq 0$,
but there is not entanglement for $A+B$ structure.

To support intuition, we express our main result on QCR in
"operational" terms: In order to use a composite system's
non-classical correlations as information theoretic resource, one
need not specifically prepare the system'a state. Rather, even if
the initial state is short of the non-classical correlations, one
can manage by targeting the alternative variables (e.g. degrees of
freedom) without any additional/intermediate operations. Some
details in this regard can be found e.g. in [12].

\bigskip

{\bf 5. Conclusion }

\bigskip

Non-classical (quantum) correlations are a matter of the composite
systems's structure, rather than that of the composite systems
itself. This Quantum Correlations Relativity is a new corollary of
quantum mechanics that is here rigorously established for the
continuous variable systems and illustrated for a typical example
for a qubits system. Physically, we realize that the quantum
information resources are ubiquitous in the bipartitions of the
composite quantum systems. From the operational perspective, our
observation suggests the quantum information resources can be
directly used without specific state preparation.

\bigskip

{\bf ACKNOWLEDGEMENTS} The authors are grateful to the referee for
very helpful suggestions. The work on this paper is financially
supported by Ministry of Science Serbia grant no 171028, and in
part for MD  by the ICTP-SEENET-MTP grant PRJ-09 ``Strings and
Cosmology``  in frame of the SEENET-MTP Network.

\bigskip

{\bf References}

\bigskip

[1] M. A. Nielsen, I. L. Chuang,  "Quantum Computation and Quantum
Information" (Cambridge Univ. Press, Cambridge, 2000).

[2] H. Ollivier, W. H. Zurek, "Quantum Discord: A Measure of the
Quantumness of Correlations" Phys. Rev. Lett. {\bf 88}, 017901
(2001).

[3]  L. Henderson, V. Vedral, "Classical, quantum and total
correlations" J.  Phys. A: Math. Gen. {\bf 34}, 6899 (2001).

[4] K. Modi, A. Brodutch, H. Cable, T. Paterek, V. Vedral,
""Quantum discord and other measures of quantum correlation""
arXiv:1112.6238v1 [quant-ph].

[5] L. C. C\' eleri, J. Maziero, R. M. Serra, "Theoretical and
experimental aspects of quantum discord and related measures" Int.
J. Qu. Inform. {\bf 9}, 1837 (2011).

[6] J.-S. Xu, C.-F. Li, "Quantum discord in open quantum systems"
arXiv:1205.0871v1 [quant-ph].

[7] S. Luo, "Quantum discord for two-qubit systems" Phys. Rev. A
{\bf 77},042303 (2008)

[8] S. Luo, "Using measurement-induced distirbance to characterize
correlations as  classical or quantum" Phys. Rev. A {\bf 77},
022301 (2008)

[9] P. Zanardi,  "Virtual Quantum Subsystems" Phys. Rev. Lett.
{\bf 87}, 077901 (2001).

[10] Dugi\' c, J. Jekni\' c,  "What is 'System': Some
Decoherence-Theory Arguments" Int. J. Theor. Phys. {\bf 45},
2249-2259 (2006).

[11] E. Ciancio, P. Giorda, P. Zanardi, "Mode transformations and
entanglement relativity in bipartite Gaussian states" Phys. Lett.
A {\bf 354}, 274-280 (2006).

[12] M. Dugi\' c, J. Jekni\' c-Dugi\' c,  "What Is 'System': The
Information- Theoretic Arguments" Int. J. Theor. Phys. {\bf 47},
805-813 (2008).

[13] A. C. De la Torre et al,  "Entanglement for all quantum
states" Europ. J. Phys. {\bf 31}, 325-332 (2010).

[14] N. L. Harshman, S. Wickramasekara, "Galilean and Dynamical
Invariance of Entanglement in Particle Scattering" Phys. Rev.
Lett. {\bf 98}, 080406 (2007).

[15] M. O. Terra Cunha, J. A. Dunningham, V. Vedral, "Entanglement
in single-particle systems" Proc. R. Soc. A {\bf 463}, 2277-2286
(2007).

[16] L. S. Schulman, "Evolution of Wave-Packet Spread under
Sequential Scattering of Particles of Unequal Mass"  Phys. Rev.
Lett. {\bf 92}, 210404 (2004)

[17] S. Gharibian, "Strong NP-Hardness of the Quantum Separability
Problem" Quantum Inf. and Comp. {\bf 10}, 343 (2010).

[18] A. Ferraro, et al, "Almost all quantum states have
non-classical correlations"  Phys. Rev. A {\bf 81}, 052318 (2010).

[19] R. M. Gingrich, C. Adami, "Quantum Entanglement of Moving
Bodies" Phys. Rev. Lett. {\bf 89}, 270402 (2002).

[20] L. Lamata, M. A. Martin-Delgado, E. Solano, "Relativity and
Lorentz Invariance of Entanglement Distillability"  Phys. Rev.
Lett. {\bf 97}, 250502 (2006).

[21] C. H. Bennett, G. Brassard, C. Cr\' epeau, R. Jozsa, A.
Peres, W. K. Wootters, "Teleporting an Unknown Quantum State via
Dual Classical and Einstein-Podolsky-Rosen Channels" Phys. Rev.
Lett. {\bf 70}, 1895-1899 (1993)

\end{document}